\documentstyle[psfig]{hep99}
\def\beq{\begin{equation}}
\def\eeq{\end{equation}}
\def\beqa{\begin{eqnarray}}
\def\eeqa{\end{eqnarray}}
\begin{document}
\title{Resummation for direct photon \\
and $W$ + jet production}
\author{Nikolaos Kidonakis}
\address{Department of Physics, Florida State University,\\ 
Tallahassee, FL 32306-4350, USA\\[3pt]
E-mail: {\tt kidonaki@sg1.hep.fsu.edu}}

\abstract{I discuss the resummation of soft gluon contributions
to direct photon and \linebreak $W$ + jet production and I
present some results for the next-to-next-to-leading order 
expansions of the resummed cross sections for these processes 
near threshold.}

\maketitle

\section{Introduction}

Direct photon production and $W$ + jet production 
are processes of great interest for a number of reasons. 
Direct photon production is important for
determinations of gluon distributions while $W$ + jet production
is relevant for estimating backgrounds to new physics.
Near threshold for the production
of the final state in these processes one can resum logarithmic 
corrections originating from soft gluon emission to all orders
in perturbative QCD. These corrections are potentially large. 
A formalism for the resummation of soft-gluon contributions at 
next-to-leading logarithmic (NLL) accuracy has been recently developed
for a variety of QCD hard scattering processes, 
including heavy quark, jet, direct photon, and $W$ + jet production
(for reviews see Refs. \cite {NK,kerkyra}). 
The resummed cross sections have been expanded up to next-to-next-to-leading 
order (NNLO), thus providing new analytical higher order predictions, 
and exhibit a substantially reduced scale dependence relative 
to next-to-leading order (NLO); this is expected on theoretical 
grounds \cite{EPIC99}.  

\section{Direct photon production}

First, we discuss resummation for direct photon production 
\cite{LOS,cmn,NK,cmnov,DIS99,NKJO} in hadronic processes
\beq
h_1+h_2 \rightarrow \gamma +X \, .
\eeq
The factorized cross section may be written as a convolution
of parton distributions $\phi$ with the partonic hard 
scattering $\hat \sigma$
\beq
\sigma_{h_1h_2\rightarrow\gamma}=
\sum_f \phi_{f_1/h_1} \otimes  
\phi_{f_2/h_2} \otimes {\hat \sigma}_{f_1f_2 \rightarrow \gamma} \, .
\eeq
Near partonic threshold $\hat \sigma$ includes large
logarithmic terms that can be resummed to all orders in perturbation 
theory.
If we take moments of the above equation, with $N$ the moment
variable, we can write the partonic cross section as 
\beqa
{\tilde{\sigma}}_{f_1f_2\rightarrow \gamma}(N) \! \!
&=& \! \! {\tilde{\phi}}_{f_1/f_1}(N)\,  {\tilde{\phi}}_{f_2/f_2}(N)\,
\hat{\sigma}_{f_1 f_2 \rightarrow \gamma}(N)
\nonumber \\ && \hspace{-17mm}
={\tilde{\psi}}_{f_1/f_1}(N)\, {\tilde{\psi}}_{f_2/f_2}(N)\, 
{\tilde{J}}(N) \, H \, {\tilde{S}}(N) \, ,
\label{sigmamomdp}
\eeqa
where in the second line of Eq. (\ref{sigmamomdp}) 
we have introduced a refactorization in terms of  new center-of-mass
parton distributions $\psi$ and a jet function $J$ that absorb the 
collinear singularities from the incoming partons and outgoing
jet \cite{GS,KS,KOS,LOS}, respectively; a soft gluon function $S$ that 
describes noncollinear soft gluon emission \cite{KS,KOS}; and a 
short-distance hard scattering function $H$.

We may then  solve for $\hat{\sigma}_{f_1 f_2\rightarrow \gamma}(N)$ 
in Eq. (\ref{sigmamomdp}). After resumming the $N$-dependence of all the
functions in the above equation, we may write the resummed 
cross section as
\beqa
{\hat{\sigma}}_{f_1 f_2 \rightarrow \gamma}(N) \! \! \! &=& \! \! \!
\exp \left \{ \sum_{i=1,2} \left [E^{(f_i)}(N_i)
+E^{(f_i)}_{\rm scale}\right]\right\}
\nonumber \\ && \hspace{-26mm} \times \;
\exp \left \{E'_{(f_J)}(N) \right\} \,
H\left(\alpha_s(\mu^2)\right) \, S\left(1, \alpha_s(S/N^2)\right) 
\nonumber \\ && \hspace{-26mm} \times \; 
\exp \left[\int_\mu^{\sqrt{S}/N} {d\mu' \over \mu'} \,
2 \, {\rm Re} \Gamma_S\left(\alpha_s(\mu'^2)\right)\right] \, .
\label{rescrosect}
\eeqa
The exponent $E^{(f_i)}$ resums the ratio $\psi/\phi$ while
$E'_{(f_J)}$ resums the $N$-dependence of the outgoing
jet \cite{KOS,LOS}. The dependence on the scale is given by
$E^{(f_i)}_{\rm scale}$. Finally, ${\rm Re} \Gamma_S$ is the
real part of the soft anomalous dimension $\Gamma_S$ which is
determined from renormalization group analysis of the soft function
$S$ \cite{KS,KOS} and has been calculated explicitly at one-loop
\cite{LOS,NK}. Because of the simpler color structure, $\Gamma_S$
is here a simple function, in contrast to heavy quark and jet production
where it is a matrix in color space \cite{KS,KOS,NKJSRV}.

Since the direct photon production cross section has been 
calculated only up to NLO,
expansions of the resummed result to higher orders provide 
new analytical predictions.
Here we present the NNLO expansion for the partonic subprocess
\beq
q(p_1)+g(p_2) \rightarrow \gamma(p_{\gamma})+q(p_J) \, .
\eeq
We define the kinematical variables $s=(p_1+p_2)^2$, 
$t=(p_1-p_{\gamma})^2$, $u=(p_2-p_{\gamma})^2$,
and $v \equiv 1+t/s$. The threshold region is given in terms of 
the variable $w \equiv -u/(s+t)$ by $w=1$. Then at $n$th order
in $\alpha_s$ we encounter plus distributions of the form
$[\ln^m(1-w)/(1-w)]_+$ with $m \le 2n-1$.

The NLO soft gluon corrections in the $\overline {\rm MS}$ scheme
for this partonic channel are given by \cite{LOS,NKJO}
\beqa
vw(1-v)s\frac{d{\hat \sigma}^{\rm NLO}_{qg 
\rightarrow q\gamma}}{dv \, dw}
&=&\sigma^B_{qg\rightarrow q\gamma} \frac{\alpha_s(\mu_R^2)}{\pi} 
\nonumber \\ && \hspace{-35mm} \times
\left\{(C_F+2C_A) \left[\frac{\ln(1-w)}{1-w}\right]_+ \right.
\nonumber \\ && \hspace{-32mm}
{}+\left[C_F\left(-\frac{3}{4}+\ln v \right)
-C_A\ln\left(\frac{1-v}{v}\right)\right.
\nonumber \\ && \hspace{-30mm} \left.
-(C_F+C_A)\ln\left(\frac{\mu_F^2}{s}\right)\right]
\left[\frac{1}{1-w}\right]_+
\nonumber \\ && \hspace{-32mm} \left.
{}+{\cal O}(\delta(1-w)) \right\} \, ,
\eeqa
with $\mu_F$ and $\mu_R$ the factorization and renormalization scales.
Here the Born term is \cite{GV}
\beq
\sigma^B_{qg\rightarrow q\gamma}=\frac{1}{N_c} \pi \alpha \alpha_s(\mu_R^2)
e_q^2 T_{qg} v \, ,
\eeq
where $T_{qg}=1+(1-v)^2$. 
Agreement is found with the exact NLO cross section in Ref. \cite{GV}.
All $\delta(1-w)$ terms can be obtained by matching to the exact NLO result.

\begin{figure}
\centerline{
\psfig{file=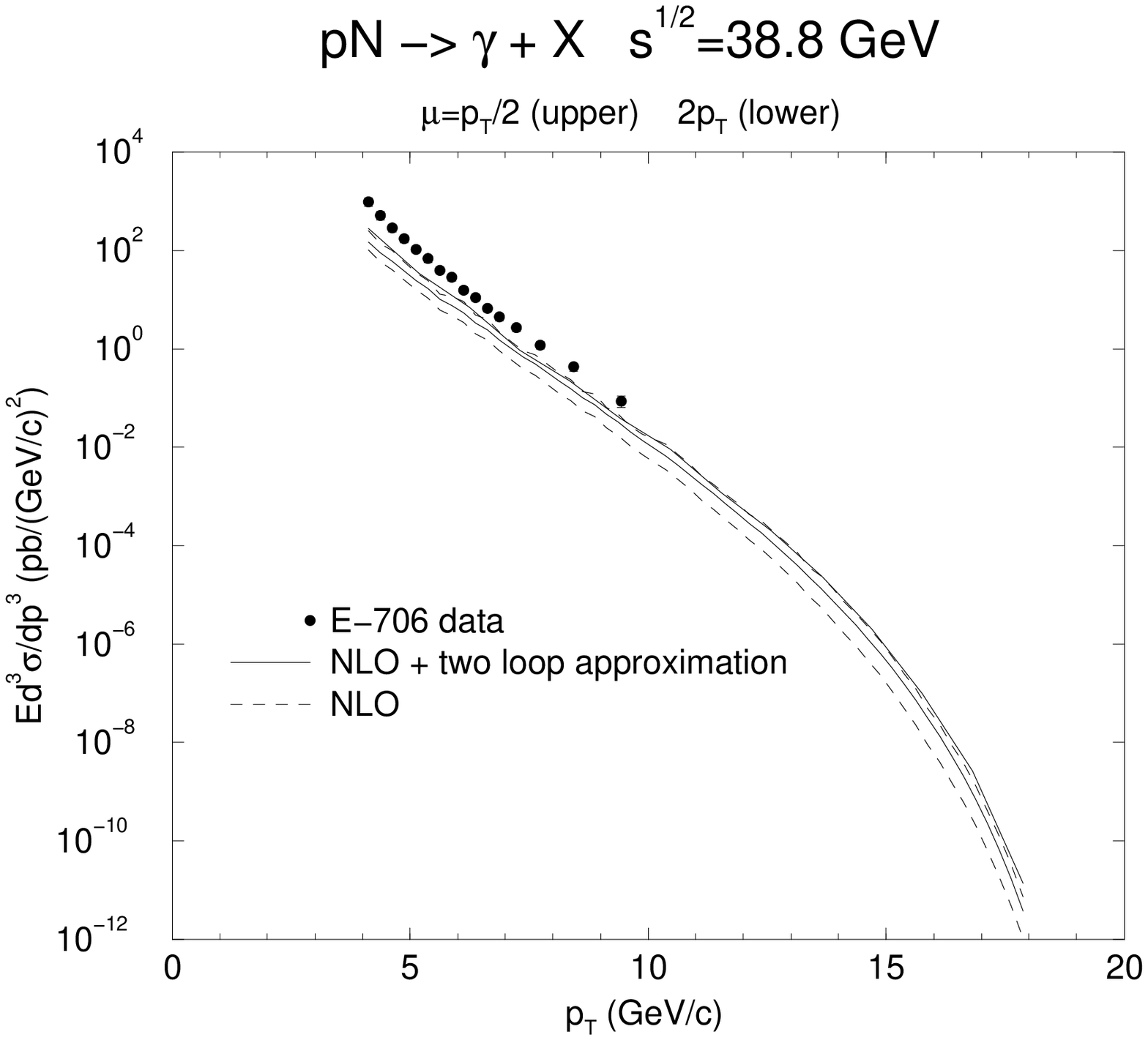,height=2.9in,width=2.9in,clip=}}
{Figure 1. Exact NLO and approximate NNLO results for 
direct photon production.}
\label{fig1}
\end{figure}

The NNLO $\overline {\rm MS}$ soft gluon corrections 
for $qg \rightarrow q\gamma$ are \cite{DIS99,NKJO} 
\beqa
vw(1-v)s\frac{d{\hat \sigma}^{\rm NNLO}_{qg \rightarrow q\gamma}}
{dv \, dw}
&=&\sigma^B_{qg\rightarrow q\gamma} \frac{\alpha_s^2(\mu_R^2)}{\pi^2} 
\nonumber \\ && \hspace{-38mm} \times \,
\left\{\frac{1}{2}(C_F+2C_A)^2
\left[\frac{\ln^3(1-w)}{1-w}\right]_{+} \right.
\nonumber \\ && \hspace{-36mm}
{}+\left[\frac{3}{2} C_F^2\left(-\frac{3}{4}+\ln v
-\ln\left(\frac{\mu_F^2}{s}\right)\right) \right.
\nonumber \\ && \hspace{-34mm}
{}+3C_A^2\left(\ln\left(\frac{v}{1-v}\right)
-\ln\left(\frac{\mu_F^2}{s}\right)\right)
\nonumber \\ && \hspace{-34mm} 
{}+\frac{3}{2}C_F C_A\left(-\ln(1-v)+3\ln v 
-3\ln\left(\frac{\mu_F^2}{s}\right)\right.
\nonumber \\ && \hspace{-34mm} \left. \left. 
{}-\frac{3}{2}\right)-\frac{\beta_0}{2}\left(\frac{C_F}{4}+C_A\right)\right]
\left[\frac{\ln^2(1-w)}{1-w}\right]_{+} 
\nonumber \\ && \hspace{-30mm} \left.
{}+{\cal O}\left(\left[\frac{\ln(1-w)}{1-w}\right]_+\right)\right\} \, ,
\eeqa
with $\beta_0=(11C_A-2n_f)/3$.
We have also derived all NNLL terms in the NNLO expansion by matching
with the exact NLO cross section \cite{NKJO}.
Analogous results have been obtained for the partonic channel
$q {\bar q}  \rightarrow g \gamma$
\cite{NK,NKJO}.

In Fig. 1 we show some numerical results \cite{NKJO}
for the direct photon production cross section as a function of
the photon $p_T$ and compare with the experimental results 
from the E706 Collaboration at Fermilab \cite{E706}.
We see that the sum of the exact NLO cross section and NNLO
approximate corrections shows a much reduced
dependence on the factorization scale relative to the exact NLO
cross section alone. However, the NNLO cross section is still  
below the E706 data.

\section{$W$ + jet production}

In this section, we discuss resummation for $W$ + jet production 
\cite{NK,NKVD} in hadronic processes
\beq
h_1+h_2 \rightarrow W +X \, .
\eeq
The construction of the resummed cross section is completely 
analogous to what we presented in the previous section for direct photon
production, Eq. (\ref{rescrosect}). The soft anomalous dimensions
are the same as for direct photon production.
In this section we expand the NLL resummed cross section for 
$W$ + jet production up to NNLO
and perform a comparison with the NLO fixed-order results of
Refs.~\cite{AR,gpw}. 
We give explicit results for the partonic subprocess 
\beq
q(p_1)+ {\bar q}(p_2) \rightarrow W(Q) + g(p_J) \, . 
\eeq
The threshold region is given in terms of the variable 
$s_2=s+t+u-Q^2$ by $s_2=0$, and we find plus distributions 
of the form $[\ln^m(s_2/Q^2)/s_2]_+$ at every order in perturbative QCD.

The NLO $\overline {\rm MS}$ soft gluon corrections for this 
partonic process at NLL accuracy are \cite{NK,NKVD}
\beqa
E_Q \frac{d{\hat\sigma}^{\rm NLO}_{q{\bar q} \rightarrow gW}}{d^3Q}&=&
\sigma^B_{q{\bar q} \rightarrow gW} {\alpha_s(\mu_R^2)\over\pi}
\nonumber \\  && \hspace{-25mm} \times \,
\left\{(4C_F-C_A) \left[\frac{\ln(s_2/Q^2)}{s_2}\right]_+  \right.
\nonumber \\  && \hspace{-22mm}
{}+\left[\left(2 C_F- C_A \right) \ln\left(\frac{s Q^2}{tu}\right) 
-\frac{\beta_0}{4} \right.
\nonumber \\  && \hspace{-20mm} \left.
{} -2 C_F \ln\left({\mu_F^2\over Q^2}\right)\right] 
\left[\frac{1}{s_2}\right]_+ 
\nonumber \\  && \hspace{-22mm} \left. 
{} + {\cal O}(\delta(s_2) \right\} \, . 
\eeqa
Here the Born term is
\beqa
\sigma^B_{q{\bar q} \rightarrow gW}&=&\frac{\alpha \alpha_s(\mu_R^2)C_F}{sN_c}
(|L_{f_2f_1}|^2+|R_{f_2f_1}|^2)
\nonumber \\ && \times 
\left(\frac{u}{t}+\frac{t}{u}+\frac{2Q^2s}{tu}\right) \, ,
\eeqa
with $L$ and $R$ the left- and right-handed couplings of the
$W$ boson to the quark line of flavor $f$ \cite{gpw}.
The NLO expansion is in agreement with the exact one-loop results in 
Ref. \cite{gpw}. All $\delta(s_2)$ terms can be obtained by matching to the 
exact NLO result.

The NNLO $\overline {\rm MS}$ soft gluon corrections for the 
partonic process $q{\bar q} \longrightarrow gW$ are \cite{NKVD}
\beqa
E_Q \frac{d{\hat\sigma}^{\rm NNLO}_{q{\bar q} \rightarrow gW}}{d^3Q}&=&
\sigma^B_{q{\bar q} \rightarrow gW}
\left({\alpha_s(\mu_R^2)\over\pi}\right)^2
\nonumber \\ &&  \hspace{-28mm} \times \,
\left\{ \frac{1}{2}(4C_F-C_A)^2\, 
\left[\frac{\ln^3(s_2/Q^2)}{s_2}\right]_+ \right.
\nonumber \\ &&  \hspace{-25mm}
{}+\left[-\frac{3}{2} (4C_F-C_A)\left(2 C_F 
\ln\left(\frac{\mu_F^2}{Q^2}\right) \right. \right.
\nonumber \\ &&  \hspace{-20mm} \left.
{}+ \left(2 C_F-C_A \right) \ln\left(\frac{tu}{s Q^2}\right) 
\right) 
\nonumber \\ && \hspace{-23mm} \left.
{}-\frac{\beta_0}{2} \left(5C_F -{3\over 2} C_A \right) \right]
\left[\frac{\ln^2(s_2/Q^2)}{s_2}\right]_+   
\nonumber \\ && \hspace{-25mm} \left.
{}+{\cal O}\left(\left[\frac{\ln(s_2/Q^2)}{s_2}\right]_+\right)\right\} \, .
\eeqa
We can also derive all NNLL terms in the NNLO expansion by matching
with the exact NLO cross section \cite{NKVD}.
Analogous results have been obtained for the partonic channel
$qg \rightarrow qW$ \cite{NK,NKVD}.
In future work we plan to study the numerical
significance of resummation for $W$+ jet production. 

This work was supported in part by the US Department of Energy.

\end{document}